\documentclass[pre,amssymb,superscriptaddress,showpacs,floatfix,a4paper,twocolumn]{revtex4-1}

\usepackage{graphicx}
\usepackage{dcolumn}
\usepackage{bm}
\usepackage{color}
\usepackage{amsmath}
\usepackage{amssymb}
\usepackage{xcolor,cancel}

\begin{document}

\title{Temperature dependence of coercivity for isolated Ni nanowires 
unraveled by high-sensitivity micromagnetometry}

\author{E. A. Escudero Bruna}
\affiliation{ 
Facultad de Ciencias Exactas y Naturales, 
Universidad Nacional de Cuyo, 
Mendoza, Argentina}
\author{F. Rom\'a}
\affiliation{
Facultad de Ciencias F\'isico, Matem\'aticas y Naturales,
Universidad Nacional de San Luis, 
Instituto de F\'isica Aplicada (INFAP), 
Consejo Nacional de Investigaciones Cient\'ificas y T\'ecnicas (CONICET), 
San Luis, Argentina}
\author{F. Meneses}
\affiliation{
Facultad de Matem\'atica, Astronom\'ia, F\'isica y Computaci\'on, 
Universidad Nacional de C\'ordoba, 
C\'ordoba, Argentina}
\author{P. G. Bercoff}
\affiliation{
Facultad de Matem\'atica, Astronom\'ia, F\'isica y Computaci\'on, 
Universidad Nacional de C\'ordoba, 
Instituto de F\'isica Enrique Gaviola (IFEG), 
Consejo Nacional de Investigaciones Cient\'ificas y T\'ecnicas (CONICET),
C\'ordoba, Argentina}
\author{M. I. Dolz}
\affiliation{
Facultad de Ciencias F\'isico, Matem\'aticas y Naturales,
Universidad Nacional de San Luis, 
Instituto de F\'isica Aplicada (INFAP), 
Consejo Nacional de Investigaciones Cient\'ificas y T\'ecnicas (CONICET), 
San Luis, Argentina}

\begin{abstract}
Magnetic nanowires (NWs) are critical components in fields such as data storage and spintronics, 
where precise control of their magnetic properties is essential for device optimization.
The behavior of isolated NWs is often 
different from that of an ensemble, offering an opportunity 
to explore the role that dipolar and magnetoelastic interactions play in the latter system.
Unfortunately, the comparison between a collection of NWs and single ones 
is often poorly characterized, as measuring individual NWs with weak magnetic signals is a challenging task. 
In this work, we employ a highly sensitive micromechanical torsional oscillator to measure 
the magnetic response of a few individual Ni NWs with
$(72 \pm 5)$ nm average diameter, 
fabricated by electrodeposition in anodic aluminum oxide templates as an array 
and subsequently released from this membrane. 
When comparing the magnetic properties as a function of temperature 
between single NWs and the array, we show that coercivity values 
of individual NWs are at least twice as large as for the array in the range $5$-$200$ K. 
Also, we characterize the differences in the hysteresis loops, 
which are more squared for isolated NWs, with a high magnetic remanence 
$\sim 80 \%$ of the saturation value. 
Our results highlight the crucial role of dipolar and mechanical 
interactions in modifying the magnetic behavior of NW arrays, 
providing valuable insights for the design and application of NW-based magnetic devices.
\end{abstract}

\maketitle

\section{Introduction}

Magnetic nanomaterials have garnered significant interest due to their singular physical properties 
and promising applications in various technical fields, including data storage, 
spintronics, and biomedical engineering
\cite{Poole2003,Cao2004,Ferry2008,Zhou2014,Liu2016,Niemann2016,Gutierrez2018}. 
Among these materials, magnetic nanowires (NWs) stand out due to their high aspect ratio, 
tunable magnetic properties, and enhanced anisotropy, making them ideal candidates 
for future next-generation technological advances \cite{Fert1999}.

Using the electrodeposition technique, ferromagnetic Ni NWs 
with a variety of sizes and morphologies 
can be synthesized inside the pores of an anodic aluminum oxide 
(AAO) template \cite{Razeeb2009,Proenca2012}.
The magnetic hysteresis properties of 
these structures have been extensively studied 
\cite{Whitney1993,Sellmyer2001,Kato2004,Kumar2006,Meneses2015,Meneses2018,Campos2020}. 
In general, within a wide range of temperatures $T$ between $5$ and $300$ K, 
a monotonic reduction in coercivity $H_c$ with increasing $T$
is observed in arrays of Ni NWs with diameters larger than $d \approx 90$ nm 
which are still embedded in the AAO template \cite{Adeela2015}. 
This is a typical behavior also found 
in bulk multidomain Ni ferromagnets \cite{Hayashi1964}, 
single-domain Ni nanoparticles \cite{Fonseca2003}, and Ni films \cite{Hanson1997}.
Interestingly, the coercivity of thinner NWs 
still embedded in the template shows
an infrequent behavior: $H_c$ increases with increasing $T$ and 
in some cases it exhibits a maximum around room temperature 
\cite{Kumar2006,Meneses2018,Campos2020,Zeng2002,Navas2008,Michel2017}.

Authors of previous studies have suggested that the internal microstructure of each wire 
(i.e., the mean grain size and the magnetocrystalline anisotropy)
and the dipolar interactions between the NWs 
are not the origin of this unexpected behavior in $H_c$.
Instead, this phenomenon could simply be due to 
the competition between the shape anisotropy of individual NWs; 
the magnetoelastic anisotropy
arising from the mechanical interaction between the NWs, the AAO template, and the Al substrate;
and the dipolar interactions between NWs \cite{Kumar2006,Meneses2018,Zeng2002}.  
Experimental data seem to confirm these conjectures.

In this work, we use a silicon micromechanical torsional oscillator
to study the hysteresis behavior of isolated Ni NWs 
in a temperature range of $5$-$200$ K.
In the absence of the template, substrate and dipolar interactions,
we measure a coercivity which depends only 
on the intrinsic characteristics of the NWs.  

The outline of the paper is as follows.  In Sec.~\ref{Exper},
we briefly describe how the Ni NWs were synthesized and released from the template, 
focusing on the most relevant characteristics of the magnetic sample we have studied.
We also describe the basic operation of the micromechanical torsional oscillator.
In Sec.~\ref{Results}, we present the results and discuss their implications.
Finally, Sec.\ref{Conclusions} is devoted to the summary and conclusions. 

\section{Experiments \label{Exper}}

\subsection{Ni NWs \label{Nistruct}}
We study Ni NWs that were synthesized in a previous work \cite{Meneses2018}.
We focus on sample DC65, where the NWs were grown 
by direct current (DC) electrodeposition inside the pores of an AAO template.
In Ref.~\cite{Meneses2018}, the magnetic hysteresis loops of this sample
were measured in a super conducting quantum interference device (SQUID),
in a temperature range of $5$-$300$ K.
The coercivity of sample DC65 
increases with increasing $T$ and 
exhibits a maximum around room temperature.
{\em A priori}, it is not clear whether this phenomenon reflects 
the magnetic properties of a single NW or rather the collective 
behavior of an array of NWs still embedded in the AAO template.

\begin{figure}[t!]
\begin{center}
\includegraphics[width=8cm,clip=true]{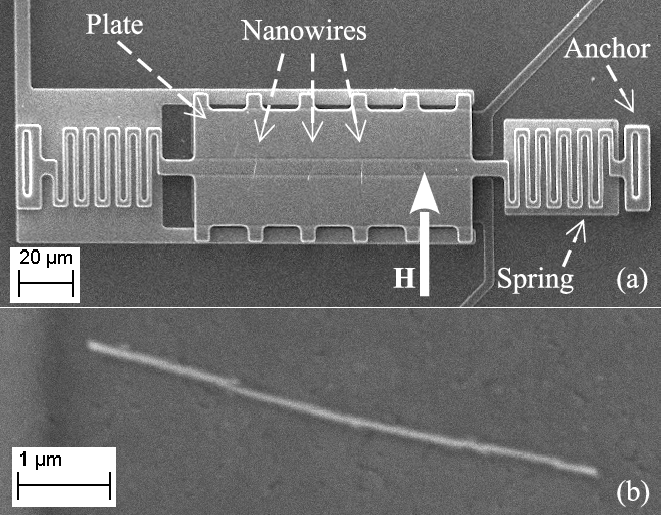}
\caption{
(a) SEM image of the silicon micromechanical oscillator 
showing the NWs stuck to its plate.  
This structure is attached to two springs which 
are in turn anchored to the substrate, 
allowing torsional oscillations forced by an 
alternating voltage signal connected to the electrodes.  
When a magnetic field $\mathbf{H}$ is applied,
the resonance frequency of the device changes, 
allowing us to infer the magnetic moment of the sample.
(b) SEM image of a typical Ni NW.}
\label{micro-osc}
\end{center}
\end{figure}

In this context, we employed a highly sensitive micromechanical torsional oscillator to measure 
the hysteresis behavior of a few individual Ni NWs.
To release the NWs from sample DC65, the array was immersed
in a mixture of H$_3$PO$_4$ ($6$ wt.$\%$) and CrO$_3$ ($1.8$ wt.$\%$) 
at $45^\circ$C for 48 hours, resulting in the total dissolution of the AAO template.
Once the NWs were released, an optical microscope and a hydraulic micromanipulator  
were used to handle them.      
Using a submicrometer drop of Apiezon N grease, a few NWs
were glued onto the plate of a silicon micromechanical oscillator.
The reason a few NWs were deposited on the microsensor 
is that they are extremely small and, if only one had been glued, 
the magnetic moment of the sample would have been too low to be measured accurately.
Figure~\ref{micro-osc}(a) shows a scanning electron microscope
(SEM) image of the measuring system. 
Since manipulation at this scale is a really difficult task, 
it is not possible to perfectly align all the NWs with each other.
In addition, in Fig.~\ref{micro-osc}(b) we show a SEM image of a typical Ni NW
where it can be observed that it is not perfectly cylindrical 
and presents small branches on the surface \cite{Meneses2018,Zeng2002}. 

\begin{figure}[t!]
\begin{center}
\includegraphics[width=8cm,clip=true]{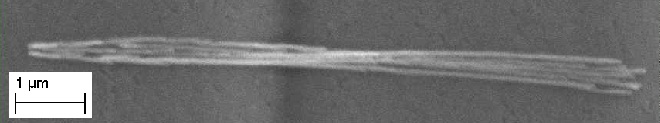}
\caption{SEM image showing a bundle made up of NWs 
of approximately $l=9$ $\mu$m in length.}
\label{bundle}
\end{center}
\end{figure}

More detailed SEM images show that some of the fragments (supposedly individual NWs)  
that were glued on the microsensor are bundles formed from just a few NWs, 
as displayed in Fig.~\ref{bundle}.
Since these clusters are made up of a few NWs, 
the dipolar interactions between them are extremely low. 
To estimate the order of magnitude of this unwanted contribution, 
we roughly calculate the longitudinal component of the magnetic field $H_z$ 
produced by a NW at a transverse distance $r$ from its center of mass, 
describing this object by two magnetic charges separated 
by a distance $l$ (the length of the NW) \cite{Kumar2006,Bertotti1998}:
\begin{equation}
H_z= \frac{m}{4 \pi \ \big(r^2+l^2/4 \big)^{3/2}}  .
\end{equation}
Here, $m=V M_b$ is the magnetic moment of a NW where $V$ and $M_b$ are, respectively,
its volume and saturation magnetization.
In this way, a single Ni NW with 
$d=72$ nm, $l=9$ $\mu$m, 
and $\mu_0 M_b=0.61$ T \cite{Bertotti1998},
where $\mu_0$ is the vacuum permeability constant, 
produces a field of approximately $\mu_0 H_z \approx 20$ $\mu$T at a distance $r=d$,
which is smaller than the magnetic field of Earth \cite{Cullity}.
This shows that the magnetic behavior of one NW belonging to a bundle 
as shown in Fig.~\ref{bundle}, will not be affected by the presence of its neighbors. 
This is true only if the number of NWs involved is very small.
In contrast, in a dense array of NWs, the long-range dipolar interaction 
is significant and therefore cannot be ignored.

In our setup, we glued 11 NWs onto the plate of the micromechanical oscillator 
and used SEM images to characterize their geometry. 
Since the NWs are branched at many points, locally enlarging their cross-sections, 
we measured an effective diameter of $(d=72 \pm 5)$ nm, 
which is slightly larger than that of the idealized cylindrical NW. 
The collection of NWs features lengths ranging from $5$ to $9$ $\mu$m, 
resulting in a total sample volume of $V_n = (3.2 \pm 0.4) \times 10^{-19}$ m$^3$.

\subsection{Micromagnetometer \label{micromag}}
The silicon micromechanical torsional oscillator 
was manufactured in the MEMSCAP Inc. Foundry \cite{MEMSCAP}.
Similar microdevices have been previously used as micromagnetometers of high
sensitivity to measure the hysteresis behavior of manganite nanotubes \cite{Dolz2008,Antonio2010,Dolz2020} 
as well as mesoscopic samples of a high-$T_c$ superconductor \cite{Dolz2007,Dolz2010}.
In simple terms, its operation can be understood as follows.
The whole system (the micromechanical oscillator with the NWs stuck on its plate) 
is cooled under vacuum inside a helium closed-cycle cryogenerator
until reaching a given temperature. 
Then the microdevice is actuated electrostatically by means of
a function generator, and its movement is sensed capacitively using a lock-in amplifier.
In this way, the resonance frequency of the system is accurately measured which depends, 
among other things, on the elastic constant of its serpentine springs.
When a uniform magnetic field (provided by an electromagnet) 
is applied along the easy axis of the NWs [see Fig.~\ref{micro-osc} (a)],
the plate of the oscillator experiences an additional torque, 
and therefore, the resonance frequency changes. 
Although extremely small, these shifts in resonance frequency ($<1$ Hz) 
can be measured and used to infer the magnetization state of the sample (see below).
Additional details of the experimental setup 
can be found in Refs.~\cite{Dolz2008,Antonio2010}. 

\begin{figure}[t!]
\begin{center}
\includegraphics[width=7cm,clip=true]{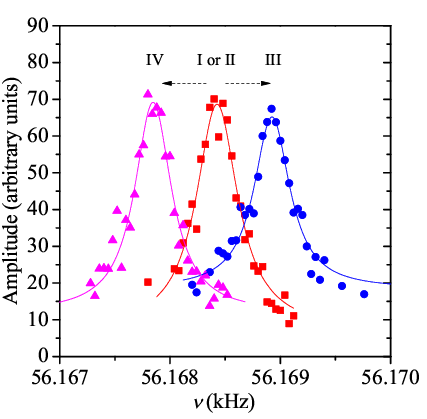}
\caption{Resonance curves obtained when $\mathbf{H}$ or $\mathbf{M}$ (or both) are zero, 
and therefore, there is no additional magnetic torque (cases I and II),
and when these fields are parallel (case III) or antiparallel (case IV),
which results in the appearance of a restoring or antirestoring magnetic torque, respectively.}
\label{curva-res}
\end{center}
\end{figure}

\begin{figure}[t!]
\begin{center}
\includegraphics[width=5.5cm,clip=true]{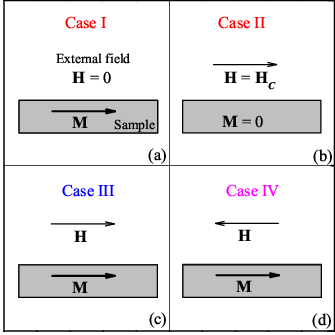}
\caption{Sketches corresponding to cases I, II, III, and IV.
No additional magnetic torque arises when (a) $\mathbf{H}=0$ or (b) $\mathbf{M}=0$ .
However, when both fields are nonzero, (c) a restoring torque appears if
$\mathbf{H}$ and $\mathbf{M}$ are parallel,
while (d) this torque is antirestoring if
$\mathbf{H}$ and $\mathbf{M}$ are antiparallel.}
\label{config}
\end{center}
\end{figure}

Typical resonance curves are plotted in Fig.~\ref{curva-res}.
With red squares, we show the amplitude of oscillation 
as a function of frequency when there is no additional magnetic torque on the system.
This happens when either no external field $\mathbf{H}$ is applied 
or when the magnetization $\mathbf{M}$ of the sample is zero,
which correspond, respectively, to cases I and II drawn in Fig.~\ref{config}
(furthermore, there is no magnetic torque in the trivial case where both vectors are zero).
When performing a hysteresis experiment, 
case II arises at the coercive field $\mathbf{H}_c$ 
since, in this instance, $\mathbf{M}=0$.   

A different situation appears when both $\mathbf{H}$ and $\mathbf{M}$ are nonzero.
As shown in case III in Fig.~\ref{config}, 
if these vectors are parallel, any slight twist with respect 
to the equilibrium position results in the appearance of a restoring magnetic torque,
which adds to the torque produced by the serpentine springs. 
As a consequence, the resonance frequency must increase 
(note that, in any mechanical oscillator, as the restoring force or torque increases, 
the resonant frequency also increases). 
The corresponding curve in Fig.~\ref{curva-res} (blue circles) 
shows this positive shift in frequency.
On the contrary, if $\mathbf{H}$ and $\mathbf{M}$ are antiparallel as in case IV,
the sample experiences an antirestoring magnetic torque,
which reduces the total torque applied to the oscillator plate, 
and then the resonance frequency decreases,
see the corresponding resonant curve in Fig.~\ref{curva-res} (magenta triangles).

The above qualitative description can be transformed into 
a quantitative formalism by making a series of simple analytical calculations 
and well-founded approximations.
Technical details are given in Appendix A.
Such calculations lead to a simple expression 
that allows the magnetization of the sample to be estimated 
from the changes in the resonance frequency $\Delta \nu$ measured in our experiments:
\begin{equation}
M=\frac{Z \Delta \nu}{H} \Bigg[ 1 + \sqrt{1+\frac{4 H^2}{Z \Delta \nu}}  \Bigg].
\label{rootMplus}
\end{equation}
Here, $H$ and $M$ are, respectively, 
the magnitudes of the external field and the magnetization, and  
\begin{equation}
Z=\frac{4 \pi^2 I \nu_0}{V_n \mu_0},
\label{constZ}
\end{equation}
where $\nu_0$ is the natural resonance frequency 
of the micromechanical torsional oscillator (when the magnetic torque is zero),
$I$ is the moment of inertia of the whole system along its center rotational axis, and   
as we defined above, $V_n$ is the volume of the magnetic sample.

\section{Results and discussion \label{Results}}

\begin{figure}[t!]
\begin{center}
\includegraphics[width=7cm,clip=true]{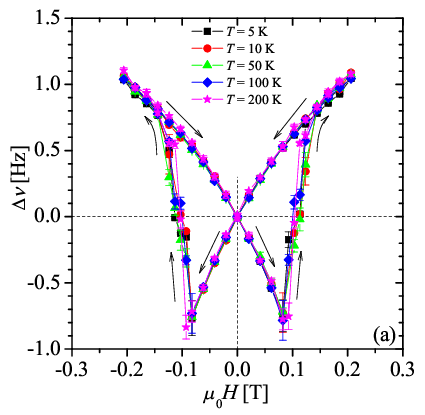}
\includegraphics[width=7cm,clip=true]{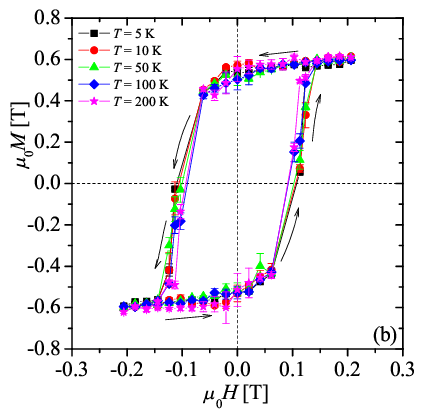}
\caption{(a) Changes in the resonant frequency $\Delta \nu$ 
as a function of the external field $\mu_0 H$ for different temperatures, as indicated. 
(b) Corresponding magnetic hysteresis loops 
calculated using Eqs.~(\ref{rootMplus}) and (\ref{constZ}).
In both figures, the arrows indicate the direction in which 
$\Delta \nu$ and $\mu_0 M$ evolve as the external field changes.}
\label{loops}
\end{center}
\end{figure}

Figure~\ref{loops}(a) shows the hysteresis loops of $\Delta \nu$ vs $H$ 
measured at temperatures $T=5$, 10, 50, 100, and 200 K.
To minimize experimental errors, 
each of these loops was obtained by averaging over at least five different measurements. 
The curves cross at $H=0$ where no changes in the resonant frequency are detected,
which corresponds to case I in Fig.~\ref{config}.
Additionally, $\Delta \nu$ is zero at the coercive field (case II).
This is a very important feature of this experimental system 
since it allows measuring the coercivity directly from
the experimental curves given in Fig.~\ref{loops}(a).
At first glance we can observe that coercivity does not depend appreciably on temperature,
being $\mu_0 H_c \sim 0.1$ T, 
a result that contrasts strongly with the observed in dense arrays of Ni NWs.

As discussed in Sec.~\ref{micromag}, 
when both the external field and the magnetization are nonzero 
and they are parallel to each other, $\Delta \nu >0$, case III in Fig.~\ref{config}.
Instead, if both vectors are antiparallel, $\Delta \nu <0$, see case IV in Fig.~\ref{config}.
This happens when the external field reverses its polarity, 
taking values between $0$ and $\pm H_c$.
The positive or negative variations of $\Delta \nu$ can be observed in Fig.~\ref{loops}(a), 
noting that the changes in the resonant frequency are, 
at most, of the order of $1$ Hz.

To calculate the magnetic hysteresis loops 
we use Eqs.~(\ref{rootMplus}) and (\ref{constZ}).  
Our micromechanical torsional oscillator has 
a resonant frequency $\nu_0 \sim 56.2$ kHz, 
a quality factor $Q$ that roughly ranges from $1.7 \times10^5$ to $1.5\times10^4$, 
and a moment of inertia along its center rotational axis 
of $I=(4.0 \pm 0.3) \times 10^{-21}$ kg m$^2$.
In Appendix B, we show how $\nu_0$, $Q$, and the sensitivity 
of this microdevice depend on temperature.

Figure~\ref{loops}(b) shows the magnetic hysteresis loops that 
we have obtained from our experimental data. 
The saturation magnetization 
does not depend appreciably on temperature and
reaches a value of $\mu_0 M_s =(0.6 \pm 0.1)$ T,
which agrees with the value for bulk Ni, $\mu_0 M_b=0.61$ T \cite{Bertotti1998}.

\begin{figure}[t!]
\begin{center}
\includegraphics[width=7cm,clip=true]{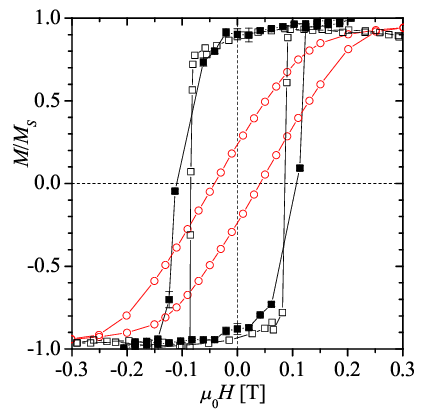}
\caption{Hysteresis loops for 
a dense array of Ni NWs still embedded in the AAO template at $T=5$K (open circles) \cite{Meneses2018},
a group of 11 isolated Ni NWs measured with 
the micromechanical oscillator at the same temperature (solid squares),
and a similar single Ni NW measured with a micro-SQUID at $T=6$ K (open squares) \cite{Wernsdorfer1996}. }
\label{loopsT5K}
\end{center}
\end{figure}

It is interesting to compare one of these magnetic hysteresis loops 
with an equivalent one (at the same temperature) previously measured for an 
array of the same Ni NWs still embedded in the alumina template \cite{Meneses2018}. 
Figure~\ref{loopsT5K} shows these curves at $T=5$ K where, 
for the sake of comparison, we have normalized the magnetization. 
The differences between them are noticeable.
The hysteresis loop of the array displays an S-shape with a low coercive field.
This is because the demagnetizing field is not negligible in a dense packing of NWs and, 
as discussed in the literature \cite{Kumar2006,Meneses2018}, 
an additional magnetoelastic effect tends to decrease 
the effective uniaxial anisotropy constant (and hence to decrease the coercivity).
In contrast, the hysteresis loop we have obtained in this work 
has a more squared shape, and the coercive field is appreciably larger.
As we have discussed previously, 
since the interactions between the NWs  
are negligible, this curve [and those shown in Fig.~\ref{loops}(b)] 
agrees with the magnetic behavior of a small group of noninteracting Ni NWs.

We also compared our measurements with the results from Ref.~\cite{Wernsdorfer1996}, 
obtained using micro-SQUID magnetometry at $T=6$ K, 
as shown in Fig.~\ref{loopsT5K} (open squares). 
There is very good agreement between both methods, 
the difference in squareness most likely related to an imperfect alignment of the NWs in our case.

\begin{figure}[t!]
\begin{center}
\includegraphics[width=7cm,clip=true]{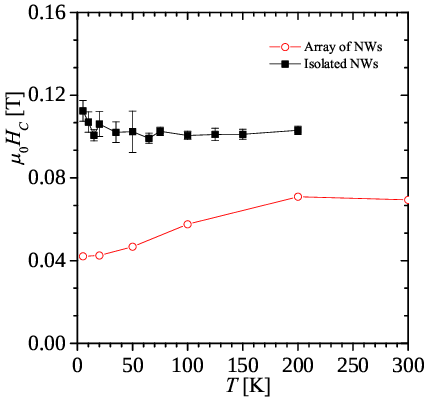}
\caption{Coercivity vs temperature for
a dense array of Ni NWs embedded in the AAO template (open circles) \cite{Meneses2018}
and the isolated Ni NWs measured with the micromechanical oscillator (solid squares).}
\label{Hc-Ms}
\end{center}
\end{figure}

Finally, in Fig.~\ref{Hc-Ms} we show the coercive field $\mu_0 H_c$ as a function of $T$
for isolated Ni NWs, determined directly from the curves given in Fig.~\ref{loops}(a).
As we observed before qualitatively,
this magnitude is nearly constant over the temperature range studied.
For comparison, it also shows the coercivity 
for an array of Ni NWs embedded in the AAO template \cite{Meneses2018},
which displays the atypical behavior mentioned throughout this work. 

The coercive field values for the isolated Ni NWs 
suggest that their magnetization is not reversed by a coherent mechanism. 
In fact, according the Stoner-Wohlfarth model \cite{Stoner1948},
the coercive field for coherent rotation at very low temperatures should be \cite{Cullity}
\begin{equation}
\mu_0 H_c = \frac{2 K_s}{M_s} = \frac{\mu_0 M_s}{2} \approx 0.3  \ \mathrm{T}, 
\end{equation}
where we have considered that $\mu_0 M_s \approx 0.6$ T for an isolated NW
and $K_s = \mu_0 M_s^2 /4$, the value of 
the shape anisotropy constant for an infinitely long rod at saturation. 
This value is at least twice as large as that measured in our experiments, 
which corroborates that other relaxation mechanisms are operating in this system.
In this regard, numerical calculations show that for small diameter Ni NWs, 
such as those studied in this work, the magnetization is reversed 
by the propagation of a Bloch-type domain wall \cite{Hertel2001}.

One approach to experimentally analyze the switching mechanisms 
is to study how the coercivity varies with the angle at which 
the external magnetic field is applied \cite{Wernsdorfer1996}. 
Unfortunately, our experimental setup makes it very difficult to explore this
problem in greater depth. 
Although, in theory, such a study could be conducted using our microdevice \cite{Kamra2014}, 
applying the external field within the plane of the oscillator plate 
causes the torque component perpendicular to the oscillation plane 
to decrease as the angle of application increases.
Consequently, the resulting change in the resonant frequency is significantly reduced 
(in fact, this torque component becomes zero when the angle reaches $90^\circ$). 
In practice, this means that coercivity can only be measured over 
a very limited angular range, making it challenging to
accurately determine the magnetization reversal mechanism.

\section{Conclusions \label{Conclusions}}

We studied the hysteresis behavior of thin ferromagnetic Ni NWs
with $(72 \pm 5)$ nm average diameter 
and lengths ranging from $5$ to $9$ $\mu$m.
After being released from the alumina template, 
11 NWs were glued 
onto the plate of a micromechanical torsional oscillator,
which was used as a micromagnetometer of high sensitivity.
Since the interactions between these NWs are negligible, 
they can be considered in effect isolated.
In this way, our measurements correspond 
to an average over this small group of noninteracting Ni NWs.

The hysteresis loops of $\Delta \nu$ vs $H$, 
obtained by applying the external field along the easy axis of the NWs,
allows direct measurement of the coercivity.    
We found that $H_c$ does not change significantly over the temperature range from $5$ to $200$ K,
which contrasts strongly with the behavior previously observed 
for dense arrays of these same Ni NWs still embedded in the AAO template.
This result confirms that the mechanical interaction between the NWs 
with both the template and the substrate
as well as the dipolar interactions between the NWs 
(which are absent in our experiments) 
is the main source of distortion of coercivity.
Also, when comparing with the Stoner-Wohlfarth model,
it is clear that the magnetization of an isolated Ni NW 
is not reversed by coherent rotation. 
 
Additionally, an analytical treatment of the data allows obtaining
the magnetic hysteresis loops of $M$ vs $H$.   
At very low temperatures, the difference between 
these curves and the corresponding loops for 
an array of Ni NWs is remarkable.

In conclusion, by using a micromechanical torsional oscillator, we have determined the differences 
in magnetic behavior between isolated NWs and the collective array. 
Our technique and analysis are powerful tools that allow exploring 
the magnetic properties of nanostructures down to the individual components, 
offering a versatile platform for advancing research across a wide variety 
of materials and applications in nanotechnology.

\section*{Acknowledgments}
This work was partially supported by CONICET under Project No. PIP 112-202001-01294-CO 
and by Universidad Nacional de San Luis under Project PROICO 03-2220 (Argentina). 
P. G. B. and F. M. acknowledge partial funding from CONICET, Foncyt-ANPCyT
and Universidad Nacional de C\'ordoba (Argentina).

\section*{Data availability}
The data that support the findings in this article are openly available \cite{raw-data}.

\appendix
\section{Magnetic hysteresis loops}

A silicon micromechanical torsional oscillator 
can be used as a high-sensitivity micromagnetometer.
The resonant frequency of the microdevice with a magnetic sample 
adhered to its plate is
\begin{equation}  
\nu_0=\frac{1}{2\pi}\sqrt{\frac{k_e}{I}}, 
\label{nu_0}
\end{equation}
where $k_e$ is the elastic restorative constant of the serpentine springs, and
$I$ is the moment of inertia of the whole system along its center rotational axis.

Because Ni NWs are ferromagnetic and have a large shape anisotropy 
well below their Curie temperature,  
an external magnetic field $\mathbf{H}$ applied parallel to the plate plane
exerts an additional restoring torque, and therefore, the new resonant frequency will be      
\begin{equation}  
\nu_r=\frac{1}{2\pi}\sqrt{\frac{k_e+k_M}{I}}. 
\label{nu_r}
\end{equation}
Here, $k_M$ is the effective elastic constant originated by the interaction 
between the magnetization of the sample $\mathbf{M}$ 
and the field $\mathbf{H}$.
As discussed in Sec.~\ref{micromag},
$k_M$ will be zero in cases I and II, positive in case III (restoring torque), 
and negative in case IV (antirestoring torque).

To calculate the magnetization of the sample, we proceed as follows.
Typically, $|\Delta \nu|=|\nu_r - \nu_0| \ll \nu_0$.
Then from Eqs.~(\ref{nu_0}) and (\ref{nu_r}), it is possible to deduce that
\begin{equation} 
k_M \simeq 8 \pi^2 I \nu_0 \Delta \nu .  
\end{equation}
This effective elastic constant depends on $\mathbf{M}$,
and it can be calculated from energy considerations \cite{Dolz2008,Zijlstra1961,Morillo1998}.
Given the high aspect ratio of the NWs,
the energy per unit volume of the sample can be written as 
\begin{equation}
U= -\mu_0 \ \mathbf{H} \cdot \mathbf{M} - \frac{K_n}{M^2} (\mathbf{M} \cdot \mathbf{n})^2,    
\end{equation}
where the first term represents the Zeeman interaction, 
and the second is the uniaxial shape anisotropy energy.
Here, $\mathbf{n}$ is a unit vector pointing along the major (easy) axis of each NW 
(parallel to the plate of the micromechanical oscillator),
and $K_n$ is the corresponding uniaxial shape anisotropy constant. 
As the oscillation amplitude is very small ($\sim 1$ sexagesimal degree at resonance),
it is possible to write that \cite{Zijlstra1961,Morillo1998} 
\begin{equation}
\frac{1}{8 \pi^2 I \nu_0 \Delta  \nu} \simeq \frac{1}{k_M} = \frac{1}{2K_nV_n}+\frac{1}{MV_n \mu_0 H},
\label{mainEq}
\end{equation}
where $V_n$ is the volume of the sample,
and we have assumed that both the temperature and the magnitude of the external field are constants. 
Equation (\ref{mainEq}) is valid if the module of magnetization 
does not change appreciably throughout the oscillation cycle,
a condition that is well satisfied given the small amplitude of oscillation.      

Taking $K_n=\mu_0 M^2 /4$, the value of 
the shape anisotropy constant for an infinitely long rod \cite{Cullity}, 
Eq.~(\ref{mainEq}) can be written as a quadratic equation in $M$
\begin{equation}
\Bigg(\frac{H}{2} \Bigg) M^2-(Z \Delta \nu) M - (2Z \Delta \nu H) = 0,
\label{EqM}
\end{equation}
where  
\begin{equation}
Z=\frac{4 \pi^2 I \nu_0}{V_n \mu_0}.
\end{equation}
The solution of Eq.~(\ref{EqM}) is 
\begin{equation}
M=\frac{Z \Delta \nu \pm \sqrt{(Z \Delta \nu)^2+4Z \Delta \nu H^2}}{H},
\label{rootM}
\end{equation}
which allows us to calculate the magnetization from the 
the experimental measurements of the change in resonance frequency $\Delta \nu$.
To guarantee that $dM/dH \ge 0$, the positive (negative) root 
of Eq.~(\ref{rootM}) should be taken when $\Delta \nu>0$ ($\Delta \nu<0$).
This rule can be avoided by rewriting Eq.~(\ref{rootM}) as
\begin{equation}
M=\frac{Z \Delta \nu}{H} \Bigg[ 1 + \sqrt{1+\frac{4 H^2}{Z \Delta \nu}}  \Bigg],
\end{equation}
where only the plus sign has been considered. 

\section{Microdevice performance}

The high sensitivity of the micromechanical oscillator is mainly due 
to the fact that the microdevice has a very high quality factor $Q$.
We define this quantity as $Q=\nu_0 / \Delta \nu_0$
where, as before, $\nu_0$ is the resonance frequency.
Here, $\Delta \nu_0$ is the bandwidth (or resonance width) 
over which the vibration power (which is proportional to 
the square of the oscillation amplitude) is greater 
than half the power at the resonant frequency
(do not mistake this quantity with $\Delta \nu$, 
which represents the change in resonant frequency 
due to the application of an external field and 
the subsequent change in the state of magnetization of the sample).    

\begin{figure}[t!]
\begin{center}
\includegraphics[width=6.9cm,clip=true]{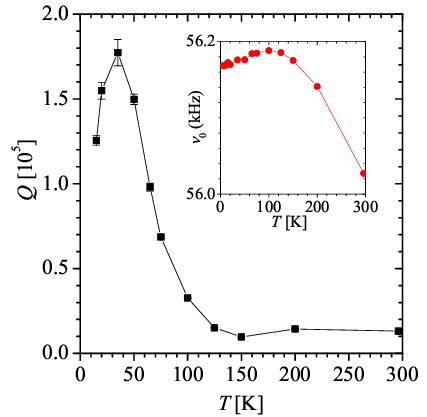}
\caption{Dependence of the quality factor $Q$ and 
the resonance frequency $\nu_0$ (inset) with temperature.}
\label{Q-fr-vsT}
\end{center}
\end{figure}

Figure~\ref{Q-fr-vsT} shows the dependence of $Q$ and $\nu_0$ (inset) with temperature.
While between $5$ and $300$ K the resonant frequency changes 
only $\sim 200$ Hz ($0.3 \%$ of its value), 
the quality factor shows more pronounced changes: 
It reaches very high values at low temperatures
but then decreases by an order of magnitude and 
stabilizes at $\sim 1.5 \times 10^4$ 
at temperatures $\ge 125$ K. 

Although the optimal performance of the microdevice evidently 
takes place in the low-temperature range, 
the values of $Q$ for $T \gtrsim 125$ K are still large enough 
to detect significant changes in the magnetization of the sample.
However, in this work, measurements with adequate precision 
could only be made up to temperatures of $200$ K.
The reason is that, although the sensitivity of the device depends on $Q$, 
it is also affected by thermal fluctuations. 

\begin{figure}[t!]
\begin{center}
\includegraphics[width=7.1cm,clip=true]{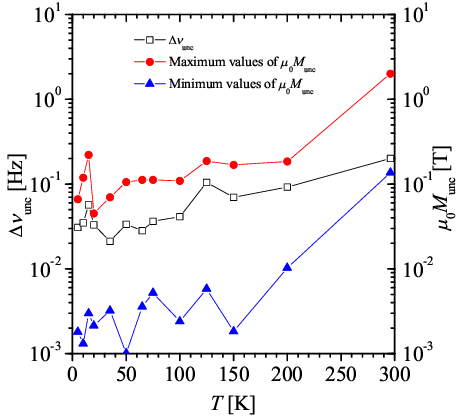}
\caption{Temperature dependence of the uncertainty values for different physical magnitudes. 
The frequency variation uncertainty $\Delta \nu_{\textrm{unc}}$ (open squares),
obtained from a direct measurement, is constant for each temperature. 
In contrast, the magnetization uncertainty $\mu_0 M_{\textrm{unc}}$,
derived indirectly, spans a range between minimum (solid triangles) and maximum (solid circles) values. }
\label{sensitivity}
\end{center}
\end{figure}

To quantify the oscillator sensitivity, 
we first considered the uncertainty of
a typical measurement of the change in 
the resonant frequency $\Delta \nu$ at each temperature $\Delta \nu_{\textrm{unc}}$.
Figure~\ref{sensitivity} shows that this quantity  
increases exponentially with $T$ and reaches a value of $0.2$ Hz at room temperature
(measurements made at $T \approx 300$ K, 
although not useful for accurately determining the coercive field, 
have been useful to estimate the sensitivity of the microdevice at this temperature).
Since $\Delta \nu_{\textrm{unc}} \approx 0.1$ Hz at $T=200$ K, 
at first glance, it would seem that by averaging over a larger number of measurements, 
it would be possible to make a sufficiently accurate estimate 
of the magnetization at room temperature.

This last observation is incorrect since our magnetometer 
does not allow direct measurement of magnetization. 
Instead, for each measurement of $\Delta \nu$, 
it is necessary to use Eq.~(\ref{rootMplus}) to calculate a single value of $M$, 
and it is these magnetization values that must ultimately be averaged. 
It is worth noting that Eq.~(\ref{rootMplus}) includes 
the $\Delta \nu/H$ and $H^2/\Delta \nu$ ratios, 
which have a significant effect on $M$ when $\Delta \nu \to 0$, i.e., 
when the applied field $H \to 0$ (case I in Fig.~\ref{config}) 
or when the coercive field is reached (case II).

We have therefore calculated $\mu_0 M_{\textrm{unc}}$, 
the uncertainty resulting from this indirect measurement of the magnetization. 
Figure~\ref{sensitivity} also shows the temperature dependence 
of the maximum (which arises for cases I and II) and minimum (for cases III and IV) 
values that this quantity takes. 
As we can see, at room temperature, the uncertainty in the magnetization 
can reach $\mu_0 M_{\textrm{unc}} \gtrsim 1$ T. 
Considering that the saturation magnetization is $\mu_0 M_s =(0.6 \pm 0.1)$ T, 
this result shows that it is not possible to obtain 
a complete magnetization loop at this temperature using our micromagnetometer.

Finally, note that between $5$ and $200$ K, 
$\mu_0 M_{\textrm{unc}}$ ranges from $\sim 10^{-3}$ to $10^{-1}$ T. 
Considering the volume of the sample, these values suggest 
that the changes in magnetic moment that can be detected 
with our oscillator range from $10^{-15}$ to $10^{-13}$ A m$^2$.



\end{document}